\documentclass{knawproc}
\usepackage{psfig}

\def\spose#1{\hbox to 0pt{#1\hss}}

\def\kms{\ifmmode {\rm\,km\,s^{-1}}\else
    ${\rm\,km\,s^{-1}}$\fi}
\def\kmsMpc{\ifmmode {\rm\,km\,s^{-1}\,Mpc^{-1}}\else
    ${\rm\,km\,s^{-1}\,Mpc^{-1}}$\fi}
\def\kpc{{\rm\,kpc}}

\def\msun{\ifmmode {\rm\,M_\odot}\else ${\rm\,M_\odot}$\fi}
\def\Msun{\ifmmode {\rm\,M_\odot}\else ${\rm\,M_\odot}$\fi}
\def\lsun{\ifmmode {\rm\,L_\odot}\else ${\rm\,L_\odot}$\fi}
\def\Lsun{\ifmmode {\rm\,L_\odot}\else ${\rm\,L_\odot}$\fi}
\def\rsun{\ifmmode {\rm\,R_\odot}\else ${\rm\,R_\odot}$\fi}
\def\Rsun{\ifmmode {\rm\,R_\odot}\else ${\rm\,R_\odot}$\fi}

\def\cm{{\rm\,cm}}
\def\cm3{\ifmmode {\rm\,cm^{-3}}\else ${\rm\,cm^{-3}}$\fi}

\def\ergps{\ifmmode {\rm\,erg\,s^{-1}}\else ${\rm\,erg\,s^{-1}}$\fi}
\def\ergpscm2{\ifmmode {\rm\,erg\,s^{-1}\,cm^{-2}}\else
    ${\rm\,erg\,s^{-1}\,cm^{-2}}$\fi}

\def\eg{{\it e.g.}}
\def\deg{\ifmmode {^{\circ}}\else {$^\circ$}\fi}
\def\degr{\ifmmode {^{\circ}}\else {$^\circ$}\fi}
\def\degs{\ifmmode {^{\circ}}\else {$^\circ$}\fi}

\def\etal{{\it et al.~}}

\def\h3Mpc{h^{-3}{\rm Mpc}^3}
\def\Ho{\ifmmode {\rm\,H_\circ}\else ${\rm\,H_\circ}$\fi}
\def\hnot{\ifmmode {\rm\,H_\circ}\else ${\rm\,H_\circ}$\fi}
\def\h0{\ifmmode {\rm\,H_\circ}\else ${\rm\,H_\circ}$\fi}
\def\hnotunit{\ifmmode {\rm\,km\,s^{-1}\,Mpc^{-1}}\else
    ${\rm\,km\,s^{-1}\,Mpc^{-1}}$\fi}
\def\lya{{\rm\,Ly-$\alpha$~}}

\def\qnot{\ifmmode {\rm\,q_\circ}\else ${\rm q_\circ}$\fi}
\def\q0{\ifmmode {\rm\,q_\circ}\else ${\rm q_\circ}$\fi}

\def\arcsec{\ifmmode {^{\prime\prime}~}\else $^{\prime\prime}~$\fi}
\def\asec{\ifmmode {^{\prime\prime}}\else $^{\prime\prime}$\fi}
\def\arcmin{\ifmmode {^{\prime}}\else $^{\prime}$\fi}
\def\amin{\ifmmode {^{\prime}}\else $^{\prime}$\fi}

\def\secper{\ifmmode \rlap.{^{s}}\else $\rlap{.}{^{s}} $\fi}
\def\minper{\ifmmode \rlap.{^{m}}\else $\rlap{.}{^m} $\fi}
\def\magper{\ifmmode \rlap.{^{m}}\else $\rlap{.}{^m} $\fi}
\def\arcsper{\ifmmode \rlap.{^{\prime\prime}}\else
    $\rlap.{^{\prime\prime}}$\fi}
\def\arcmper{\ifmmode \rlap.{^{\prime}}\else
    $\rlap.{^{\prime}}$\fi}
\def\spose#1{\hbox to 0pt{#1\hss}}
\def\simlt{\mathrel{\spose{\lower 3pt\hbox{$\mathchar"218$}}
     \raise 2.0pt\hbox{$\mathchar"13C$}}}
\def\simgt{\mathrel{\spose{\lower 3pt\hbox{$\mathchar"218$}}
     \raise 2.0pt\hbox{$\mathchar"13E$}}}

\def\aj{{AJ}}
\def\apj{{ApJ}}

\def\apjl{{ApJ}}

\def\nature{{Nature}}

\def\apjref#1;#2;#3;#4 {\par\pp#1, {#2}, #3, #4 \par}

\begin{document}

% The ``opening'' environment takes care of title, author and headlines
\begin{opening}

\title{Induced star formation and morphological evolution in very
high redshift radio galaxies}

\author{Wil van Breugel$^1$, A. Stanford$^1$\\
A. Dey$^2$, G. Miley$^3$\\ 
D. Stern$^4$, H. Spinrad$^4$, J. Graham$^4$\\
P. McCarthy$^5$\\
}
\addresses{%
1. Institute of Geophysics \& Planetary Physics, LLNL, USA\\
2. The Johns Hopkins University, USA\\
3. Leiden Observatory, The Netherlands\\
4. Astronomy Department, University of California, Berkeley, USA\\
5. Observatories of the Carnegie Institution of Washington, USA\\ 
}

\runningtitle{very high redshift  radio galaxies}
\runningauthor{van Breugel et al.}

\end{opening}

%%%%%%%%%%%%%%%%%%%%%%%%%%%%%%%%%%%%%%%%%%%%%%%%%%%%%%%%%%%%%%%%%%%%%%%%%%%%%%%

\begin{abstract}

Near-infrared, sub-arcsecond seeing images obtained with the
W.\ M.\ Keck I Telescope of show strong evolution at {\it rest--frame
optical} wavelengths in the morphologies of high redshift radio
galaxies (HzRGs) with $1.9 < z < 4.4$.  The structures change from
large--scale low surface brightness regions surrounding bright,
multiple component and often radio--aligned features at $z > 3$,  to
much more compact and symmetrical shapes at $z < 3$.  The linear sizes
($\sim 10$ \kpc) and luminosities ($M_B \sim -20$ to $-22$)
of the {\it individual} components in the $z > 3$ HzRGs are similar to
the {\it total} sizes and luminosities of normal, radio--quiet, star
forming galaxies seen at $z = 3 - 4$.

`R'-band, 0.1\arcsec resolution images with the Hubble Space Telescope
of one of these HzRGs, 4C41.17 at $z = 3.800$, show that at {\it
rest--frame UV} wavelengths the galaxy morphology breaks up in even
smaller, $\sim$ 1 \kpc--sized components embedded in a large halo of
low surface brightness emission.  The brightest UV emission is from a
radio--aligned, edge-brightened feature (4C41.17-North) downstream from
a bright radio knot.  A narrow--band \lya image, also obtained with
HST, shows an arc--shaped \lya feature at this same location,
suggestive of a strong jet/cloud collision.

Deep spectropolarimetric observations with the W.\ M.\ Keck II
Telescope of 4C41.17 show that the radio--aligned UV continuum is
unpolarized.  Instead the total light spectrum shows absorption lines
and P-Cygni type features that are similar to the radio--quiet $z = 3 -
4$ star forming galaxies.  This shows that the rest--frame UV light in
4C41.17 is dominated by starlight, not scattered light from a hidden
AGN.  The combined HST and Keck data suggest that the radio--aligned
rest--frame UV continuum is probably caused by jet--induced star
formation.

The strong morphological evolution suggests that we see the first
evidence for the assemblage of massive ellipticals, the parent
population of very powerful radio sources at much lower redshifts.  The
presence of radio--aligned features in many of the $z > 3$ HzRGs
suggests, by analogy to 4C41.17, that jet--induced star formation may
be a common phenomenon in these galaxies in their early stages of
formation.

\end{abstract}

%%%%%%%%%%%%%%%%%%%%%%%%%%%%%%%%%%%%%%%%%%%%%%%%%%%%%%%%%%%%%%%%%%%%%%%%%%%%%%%

\section{Introduction}

Radio sources have allowed the identification of luminous galaxies out
to extremely high redshifts.  Optical/near--IR campaigns during the past
few years by several groups have resulted in the discovery of numerous
radio galaxies at $z > 2$, including 20 with $z > 3$, and 3 with $z > 4$
(see De Breuck \etal 1998, this volume).  At lower redshifts powerful
radio sources are consistently identified with giant elliptical and cD
galaxies, suggesting that at high redshifts we may be observing these
massive galaxies in their early stages of evolution. While recently
developed techniques of finding very distant star--forming galaxies
are yielding substantial `normal' galaxy populations at $z \simgt 3$
(Steidel \etal 1996; Dickinson 1998), radio galaxy samples remain the
best means of finding, and studying, the most massive galaxies at the
highest redshifts.  In `traditional' models such galaxies are thought
to form early ($z_F > 5$), while in `hierarchical models this process
is thought to take much longer (\eg \ Kauffmann and Charlot 1998).
Observations of HzRGs may therefore provide a unique opportunity to
study massive forming galaxies at the highest redshifts and may help
discriminate between these two very different cosmogonies.

We discuss the results of our near--IR imaging program with the
W.\ M.\ Keck Telescope I to investigate the morphological evolution of
HzRGs between $1.9 < z < 4.4$, and a detailed study of one of these
objects (4C41.17 at $z = 3.800$) to determine the nature of its
radio--aligned continuum using the Hubble Space Telescope (continuum
and emission line imaging) and the W.\ M.\ Keck Telescope II
(spectroplarimetry).

We adopt H$_0 = 50$ km s$^{-1}$ Mpc$^{-1}$, q$_0 = 0.0$, and $\Lambda =
0$.  The assumed cosmology implies a angular size scale of 13$-$14 kpc
arcsec$^{-1}$ for the redshift range $z = 1.9 - 4.4$ covered.  For q$_0
= 0.1$ the corresponding angular size scale is 14\% -- 33\% less at
these redshifts.

\section{Keck near-infrared imaging of HzRGs}

Near-infrared, 0.4\arcsec - 0.7\arcsec seeing images obtained with the
W.\ M.\ Keck I Telescope of HzRGs with $1.9 < z < 4.4$ show strong
morphological evolution at {\it rest--frame optical} ($\lambda_{\rm
rest} > 4000$\AA) wavelengths (van Breugel \etal 1998; Figure 1).
At the highest
redshifts, $z > 3$, the rest--frame visual morphologies exhibit
structure on at least two different scales:  relatively bright, compact
components with typical sizes of $\sim$10 \kpc~surrounded by
large--scale ($\sim$ 50 $-$ 100 \kpc) diffuse emission. The brightest
components are often aligned with the radio sources, and their {\it
individual} luminosities are $M_B \sim -20$ to $-22.$ For comparison,
present--epoch L$_\star$ galaxies and, perhaps more appropriately,
ultraluminous infrared starburst galaxies, have, on average, $M_B \sim
-21.0$. The {\it total, integrated} rest--frame B--band luminosities
are $3 - 5$ magnitudes more luminous than present epoch $L_\star$
galaxies.

The presence of radio--aligned optical 
features suggests a causal connection
with the AGN.  The most popular explanations for such an alignment
effect, which is most prominent at rest--frame UV wavelengths, include
induced star formation, nebular continuum emission, or scattered light
from a hidden quasar.  At $z \sim 1$ and $z \sim 2.5$ (Dey, this
volume; Cimatti \etal, this volume) the presence of strongly polarized
light is evidence in support of the latter. However, deep Keck
spectropolarimetry of two $z > 3$ radio galaxies shows that at these
higher redshifts the light may be dominated by hot young stars 
(see section below; and Dey, this volume).

\begin{figure}
\centerline{
\psfig{figure=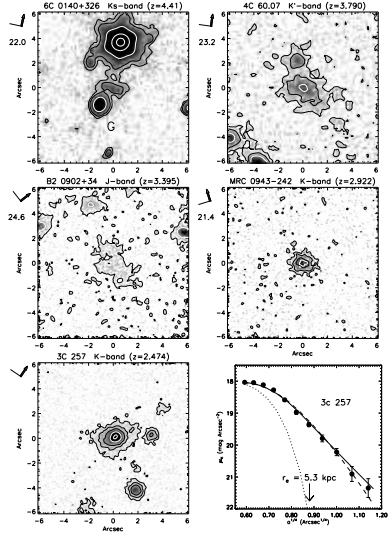,height=7.0cm}
}
\caption{\label{fig}
\small{Selected near--IR images of HzRGs, presented in order of decreasing
redshift, and the surface brightness profile of 3C257.
}}
\end{figure}
At lower redshifts, $z < 3$, the rest--frame optical morphologies
become smaller, more centrally concentrated, and less aligned
with the radio structure.  Galaxy surface brightness profiles for the
$z < 3$ HzRGs are much steeper than those of at $z > 3$.  We attempted
to fit the $z < 3$ surface brightness profiles with a de Vaucouleurs
r$^{1/4}$ law and with an exponential law, the forms commonly used to
fit elliptical and spiral galaxy profiles, respectively.  We
demonstrate the fitting for our best resolved object at $z < 3$, 3C 257
at $z = 2.474$ (Figure 1).  Within the limited dynamical range of the
data, both functional forms fit the observed profiles---neither is
preferred.  Interestingly, despite this strong morphologicsl evolution
the $K - z$ `Hubble' diagram for the most luminous radio galaxies
remains valid even at the highest redshifts, where a large fraction of
the K-band continuum is due to a radio--aligned component (see De
Breuck \etal, this volume).

\section{HST imaging of 4C 41.17}

4C41.17 at $z = 3.800$ was one of the first truly HzRGs discovered
(Chambers \etal 1990). Previous observations with the aberrated HST
showed that the optical continuum of 4C41.17 is very clumpy and aligned
with the inner radio source (Miley \etal 1992).  Much improved
observations were subsequently obtained with the refurbished HST,
including a deep rest--frame UV image (F702W filter, $\lambda{_{rest}}$
$\sim$ 1430 \AA; 6.0 hrs exposure), and a \lya image (LRF filter at
$\lambda_c \sim$ 5830 \AA; 2.0 hrs exposure).  One of the field objects
in the HST images was also seen at near--IR (Graham \etal 1994; object
\# 16) and radio wavelengths (Carilli \etal 1994).  This was used to
align the HST and radio frames with an estimated relative accuracy of
$\sim$ 0.1\arcsec.  The central, radio--aligned UV and \lya emission is
shown in Figure 2 with the 0.21\arcsec resolution radio X-band image
from Carilli \etal overlaid.  Figure 3 shows the HST continuum image of
the entire 4C41.17 system smoothed to 0.3\arcsec resolution to enhance
low surface brightness features. The star formation rates ($S.F.R.$) of
the various components as deduced from the rest--frame UV HST
photometry are listed in Table 1.  We have assumed L$_{1500\AA} \sim
10^{40.1}$ \ergps\AA$^{-1}$ for a $S.F.R.$ = 1 \msun/yr (Conti \etal
1996) and no dust reddening (which is surely a lower limit, given the
detection of dust at sub---mm wavelengths in 4C41.17 by Dunlop \etal
1994).
\tabskip=1em plus2em minus.5em
$$\vbox{\rm\offinterlineskip  \halign { \strut #\hfil &\hfil# & \hfil#\hfil & \hfil#\hfil & \hfil#\hfil & \hfil#\hfil & \hfil#\hfil & \hfil#\hfil & \hfil#\hfil\cr
\multispan{5} \hfil{Table 1} \hfil\cr
\noalign{\vskip4pt}
\multispan{5} \hfil{Luminosities and Star Formation Rates} \hfil\cr
\noalign{\vskip2pt}
\multispan{5}\hrulefill \cr
\noalign{\vskip2pt}
\multispan{5}\hrulefill\cr
\noalign{\vskip2pt}
4C41.17 & Diam. & $L_{Ly-\alpha ,c}$ & L$_{\lambda{em,c}}$ & $S.F.R$ \cr
\noalign{\vskip2pt}
Comp.   & \kpc & 10$^{43}$ \ergps & 10$^{41}$ \ergps\AA$^{-1}$ & \msun/yr \cr
\noalign{\vskip2pt}
\multispan{5}\hrulefill\cr
\noalign{\vskip2pt}
NW      & 11 &  8.3 $\pm$ 0.6 &  5.7 $\pm$ 0.2 &  60 \cr 
NE      & 11 & 21.7 $\pm$ 0.7 & 19.8 $\pm$ 0.3 & 200 \cr
NEE     & 11 & 12.4 $\pm$ 0.6 &  2.8 $\pm$ 0.2 &  30 \cr
S       & 22 & $<$ 3.8 (3$\sigma$) & 11.3 $\pm$ 0.3 & 110 \cr
Total   & 65 & 137  $\pm$ 6   & 65.9 $\pm$ 1.1 & 660 \cr
\multispan{5}\hrulefill\cr
\noalign{\vskip2pt}
N-Knots  & 3 & -- & 0.5 -- 4.0  & 5 -- 40 \cr
S-Knots  & 3 & -- & 0.5 -- 1.0  & 5 -- 10 \cr
\multispan{5}\hrulefill\cr
%\noalign{\vskip4pt}
%\multispan{5}{$^a$ Assuming L$_{1500\AA} \sim 10^{40.1}$ \ergps\AA$^{-1}$ 
%for a $S.F.R.$ = 1 \msun/yr (Conti \etal 1996)}\hfil\cr
}}$$

\subsection {Aligned UV and \lya emission: jet-induced star formation?}
\begin{figure}
\centerline{
\psfig{figure=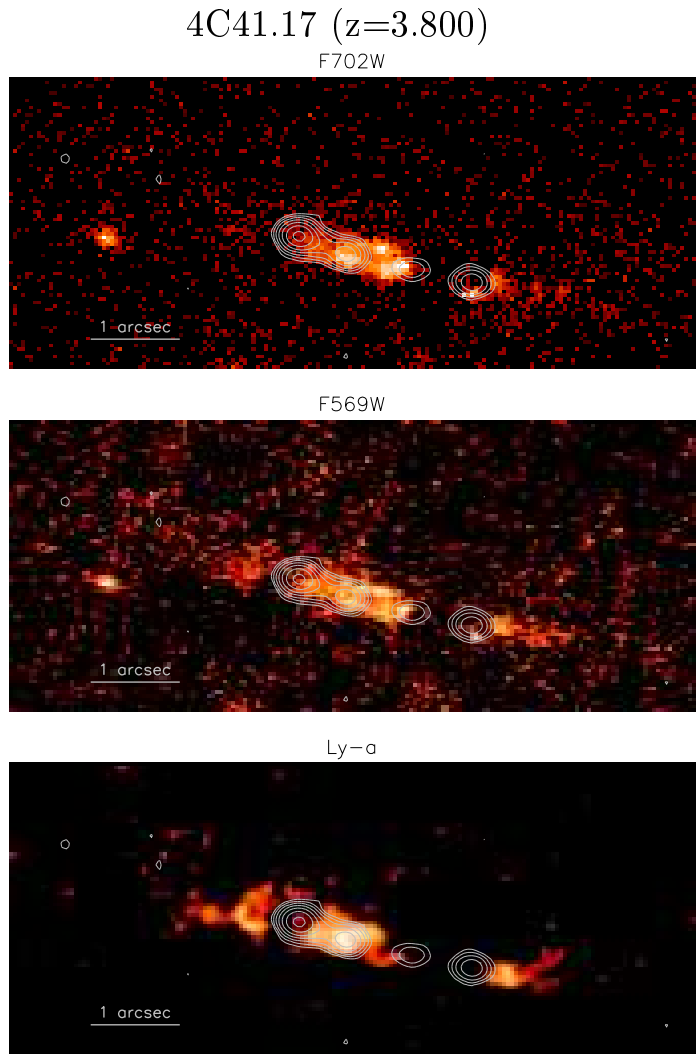,height=16cm}
}
\caption{\label{fig}
\small{HST WFPC2 images of 4C41.17-North through the F702W (`R'-band), 
F569W (includes \lya) and \lya filters. 
A X-band radio map from Carilli \etal 
is superimposed.
}}
\end{figure}
The HST images show that the 4C41.17 system consists of two
components:  4C41.17-North with a bright string of UV knots and \lya
emission along the radio axis, and 4C41.17-South with several much
fainter knots, distributed in random fashion throughout a low surface
brightness halo.  The brightest radio knot in 4C41.17-North is
associated with the brightest UV knot and arc-like \lya emission.
Downstream from this knot (NEE, Table 1) the radio source curves towards
a faint, very steep spectrum NE lobe (see Carilli \etal 1994), while
upstream (NE, Table 1) from this knot, towards the central radio
core, the UV continuum appears edge--brightened. These morphological
features suggest a strong interaction between the radio jet and dense
ambient gas and, in fact, are as expected in jet--induced star
formation models where sideways shocks induce starformation in the
dense medium of forming galaxies (\eg De Young 1989; Bicknell \etal
1998).

\subsection {The clumpy UV and \lya halo: a forming massive galaxy?}

The entire 4C41.17 system is embedded in a common halo of diffuse, low
surface brightness emission which extends over a very large area of
54$h_{50}^{-1}$~kpc $\times$ 76$h_{50}^{-1}$~kpc (5\arcsec $\times$
7\arcsec). This includes a faint region, 4C41.17-South, with half a
dozen compact knots distributed in random fashion. Spectroscopic
observations have shown that 4C41.17-South is indeed at the same
redshift as 4C41.17-North (Dey \etal 1999).  The range of UV
luminosities and $S.F.R.$ rates for the individual knots in
4C41.17-South is lower than in 4C41.17-North, and very similar to the
`normal' (radio-quiet) Lyman-break galaxies discovered by Steidel \etal
(1996). The random distribution and on average lower $S.F.R.$ in the
4C41.17-South knots suggests that star formation here is unaided by
bowshocks from the radio jet.

The Keck near-IR images have shown that the entire 4C41.17 system
(North + South) has a very blue (line-free) continuum, with $R - K_s =
2.7 \pm 0.1$ in a 3\arcsec circular aperture (Graham \etal 1994). From
these same data we have estimated that 4C41.17-South itself may even be
bluer, with $R - K_s \sim 2.2 \pm 0.2$ in a 1.9\arcsec circular
aperture.  Such colors are commonly found for galaxies in the $3 < z <
4$ range, and are thought to indicate ongoing star formation (see for
example Eisenhardt and Dickinson 1992 [B2 0902+34]; Steidel \etal 1996
[`Ly-break' galaxies]).  In 4C41.17-South perhaps as much as 80\% of
the continuum is due to low surface brightness emission so that, on the
basis of its blue colors, it appears that star formation occurs not
only in compact, kpc-sized starbursts, but also throughout the
4C41.17-South region, and indeed perhaps the entire 4C41.17 system.
The total integrated UV luminosity of 4C41.17, over a region
approximately 60\kpc~ in diameter and including the compact and low
surface brightness regions, then implies a total SFR of at least $\sim$
660 \msun/yr (Table 1). Of this perhaps as much as 2/3 of the star
formation may be occuring in the inter-knot regions. If the total star
formation would continue at this rate for $2 \times 10^{8} - 2 \times
10^{9}$ yrs an entire massive elliptical galaxy of $10^{11} \msun -
10^{12} \msun$ might be assembled between $z \sim 4$ and $z \sim 2.5$,
consistent with the morphological evolution for HzRGs seen in the
near--IR Keck observations (Section 2).
\begin{figure}
\centerline{
\psfig{figure=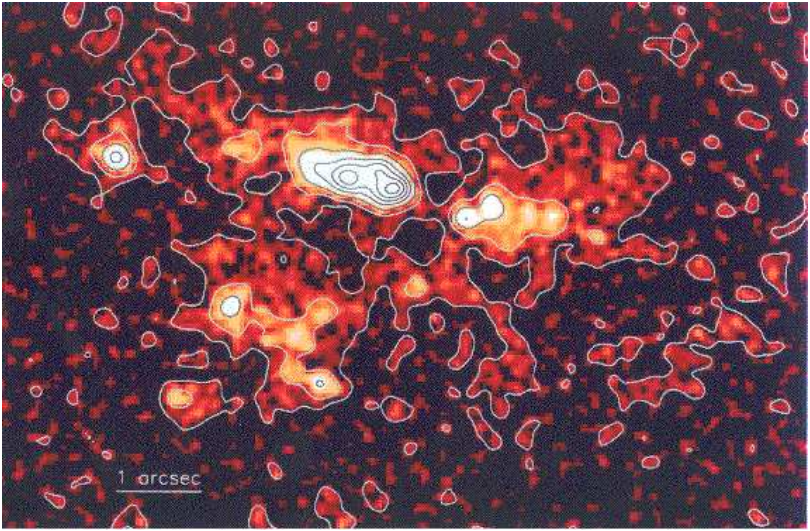,height=7.0cm}
}
\caption{\label{fig}
\small{Smoothed version of the F702W image in Figure 1 showing the clumpy companion system 4C41.17-South.
}}
\end{figure}

The HST observations of 4C 41.17 thus seem to suggest that the
formation of massive ellipticals might occur on two different length
scales.  One in which relatively dense, $1 - 10$ kpc-sized star forming
regions evolve separately and then merge, and one which involves star
formation on a very much larger ($\sim 70$ \kpc) scale.  The latter may
be more dissipative in nature, as suggested by huge ($\sim$ 200 kpc),
turbulent and possibly slowly rotating \lya halo centered on 4C 41.17
(Chambers \etal 1990; Dey \etal 1999).  Quite possibly the star
formation on these small and large scales proceeds on different time
scales and with different mass functions.  As earlier suggested by
Kormendy (1989), hybrid galaxy formation models, involving both
dissipationless merging of stellar systems and of dissipative collapse,
may be needed to understand the formation of very massive galaxies.

\section{Keck spectropolarimetry of 4C41.17}

The deep spectropolarimetric observations with the Keck II telescope by
Dey \etal (1997) have provided strong evidence in support of the
jet--induced star formation model for 4C41.17-North suggested above on
the basis of the HST and radio morphologies. These observations showed
that the bright, radio--aligned rest--frame UV continuum is unpolarized
($P_{UV}(2\sigma) < 4\%$).  This implies that scattered AGN light,
which is generally the dominant contributor to the rest-frame UV
emission in $z\sim 1$ radio galaxies, is unlikely to be a major
component of the UV flux from 4C 41.17. Instead, the total light
spectrum shows absorption lines and P--Cygni--like features that are
similar to those detected in the spectra of the recently discovered
population of star forming galaxies at slightly lower ($z\sim2-3$)
redshifts (Fig 4).  The detection of the S V$\lambda 1502$ stellar
photospheric absorption line, the shape of the blue wing of the Si IV
profile, the unpolarized continuum emission, the inability of other
AGN--related processes to account for the UV continuum flux, and the
overall similarity of the UV continuum spectra of 4C 41.17 and the
nearby star forming region NGC 1741B strongly suggest that the light
from 4C 41.17 is dominated by young, hot stars.  The presence of
radio--aligned features in many of the $z > 3$ HzRGs suggests, by
analogy to 4C41.17, that jet--induced star formation may be a common
phenomenon at these very high redshifts.

\begin{figure}
\centerline{
\psfig{figure=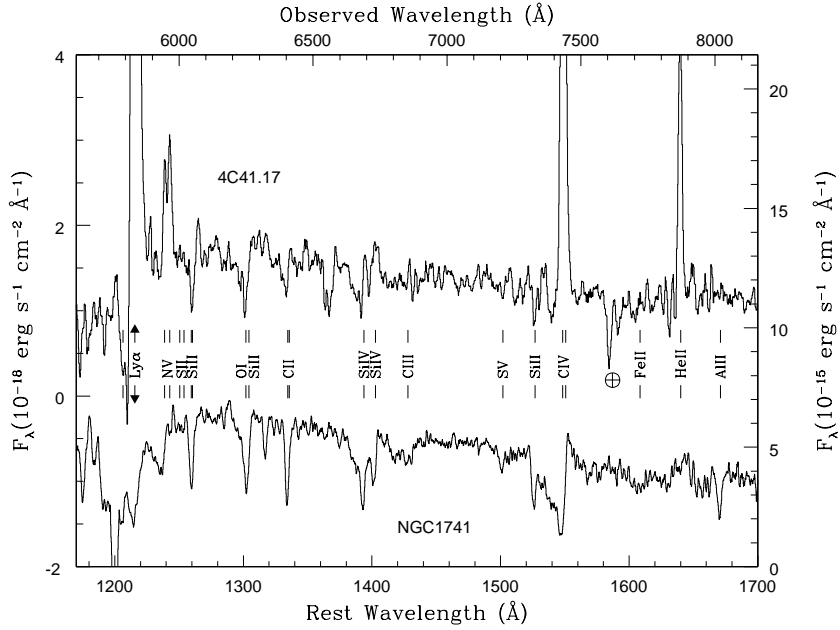,height=9.2cm}
}
\caption{\label{fig}
\small{Keck spectrum (Dey \etal 1997) of the radio-aligned component
4C41.17-North, compared with a UV spectrum of the Wolf-Rayet starburst
galaxy NGC 1741 (Conti \etal 1996).
}}
\end{figure}

\noindent {\bf Acknowledgments} 

We thank G.\ Bicknell and M.\ Dopita for stimulating discussions about
jet--induced star formation models, S.\ Rawlings for advance
Information regarding 6C~0140+326 and 8C~1435+635 and his work on the
$K - z$ diagram, and C.\ Carilli for providing high quality radio
images of 4C41.17 which allowed to improve on the relative
radio/optical astrometry for this source. The research at IGPP/LLNL is
performed under the auspices of the US Department of Energy under
contract W--7405--ENG--48.

\end{document}